\documentstyle[12pt,aasms4,epsf]{article}

\lefthead{Ogilvie \& Livio} \righthead{MHD outflows from accretion
disks}

\begin{document}

\title{On the difficulty of launching an outflow from an accretion disk}

\author{Gordon~I.~Ogilvie\altaffilmark{1} and
Mario~Livio\altaffilmark{2}} \affil{Space Telescope Science Institute,
3700 San Martin Drive, Baltimore, MD~21218} \altaffiltext{1}{Permanent
address: Institute of Astronomy, University of Cambridge, Madingley
Road, Cambridge CB3~0HA, United Kingdom.  e-mail address:
gogilvie@ast.cam.ac.uk} \altaffiltext{2}{e-mail address:
mlivio@stsci.edu}

\slugcomment{Accepted for publication in the {\it Astrophysical Journal}}

\begin{abstract}
We solve for the local vertical structure of a thin accretion disk
threaded by a poloidal magnetic field.  The angular velocity deviates
from the Keplerian value as a result of the radial Lorentz force, but
is constant on magnetic surfaces.  Angular momentum transport and
energy dissipation in the disk are parametrized by an
$\alpha$-prescription, and a Kramers opacity law is assumed to hold.
We also determine the stability of the equilibria with respect to the
magnetorotational (or Balbus--Hawley) instability.  If the magnetic
field is sufficiently strong, stable equilibria can be found in which
the angle of inclination, $i$, of the magnetic field to the vertical
at the surface of the disk has any value in the range $0\le i<90^\circ$.
By analyzing the dynamics of a transonic outflow in the corona of the
disk, we show that a certain potential difference must be overcome
even when $i>30^\circ$.  We determine this potential difference as a
function of $i$ for increasing values of the vertical magnetic field
strength.  For magnetorotationally stable equilibria, the potential
difference increases faster than the fourth power of the magnetic
field strength, quickly exceeding a value corresponding to the central
temperature of the disk, and is minimized with respect to $i$ at
$i\approx38^\circ$.  We show that this property is relatively
insensitive to the form of the opacity law.  Our results suggest that
an additional source of energy, such as coronal heating, may be
required for the launching of an outflow from a magnetized disk.
\end{abstract}


\section{Introduction}

The highly collimated outflows associated with many astrophysical
objects, from young stellar objects to active galactic nuclei, are
widely believed to originate in accretion disks threaded by magnetic
fields (see, e.g., Livio\markcite{L97} 1997 for a recent review).
While winds from rotating stars such as the Sun are influenced by
magnetohydrodynamic (MHD) effects, in particular with regard to the
magnetic braking of the star (e.g.~Mestel\markcite{M68} 1968), these
outflows are driven predominantly by thermal effects in the corona of
the star.  In the case of accretion disks, however, the possibility
arises of a wind driven predominantly by dynamical effects.

The influential model of Blandford \& Payne\markcite{BP82} (1982)
established the significance of the angle of inclination, $i$, of the
poloidal magnetic field lines to the vertical at the surface of the
disk.  They showed that, when $i>30^\circ$, an outflow is driven
spontaneously by the centrifugal force (as viewed in the frame of
reference rotating at the local Keplerian angular velocity).  This
effect occurs because, for material forced to rotate at the angular
velocity of the foot-point of the magnetic field line, the
centrifugal--gravitational potential decreases, rather than increases,
along the field line leaving the surface of the disk.

This result should not be interpreted as meaning that a wind can flow
freely and steadily from the surface of the disk without any thermal
assistance.  If that were possible, it would lead to the dynamical
disruption of the entire disk, and it would be more correct to say
that the disk was never in equilibrium to begin with.  Blandford \&
Payne\markcite{BP82} (1982) appreciated that their model of a `cold'
wind does not properly describe the flow in the neighborhood of the
surface of the disk, where thermal effects must become important.
Indeed, the flow must pass through a sonic (strictly, slow
magnetosonic) point in this region and can do so only with thermal
assistance.  The purpose of our investigation is to determine
quantitatively how much assistance is required.

We are interested in the internal equilibrium of the disk and in the
properties of the wind in the region immediately above the surface of
the disk.  It is therefore sufficient to consider a radially localized
region of the disk and solve for its vertical structure.  The present
work is most closely related to that of Wardle \&
K\"onigl\markcite{WK93} (1993), who applied a similar procedure to
weakly ionized, protostellar disks in which ambipolar diffusion is
important.  There are several significant differences in our approach,
however, which are described in detail at the end of this paper.

In several other studies (e.g.~Cao \& Spruit\markcite{CS94} 1994;
Lubow, Papaloizou, \& Pringle\markcite{LPP94} 1994; Kudoh \&
Shibata\markcite{KS97} 1997), the shape and angular velocity of the
magnetic field lines is assumed, but we determine them
self-consistently by solving the relevant equations.  Another approach
used recently (e.g.~Ouyed \& Pudritz\markcite{OP97} 1997) is to
solve for the wind flow in the corona of the disk using axisymmetric
numerical simulations.  We emphasize that, since these simulations
describe the flow only {\it after} it has become supersonic, they are
complementary to, but quite distinct from, the present work.

Recently, Livio\markcite{L97} (1997) reviewed the observations of all
the classes of astrophysical objects with which jets are associated.
Working on the assumption that the same mechanism of acceleration and
collimation operates in all these systems, he argued that the most
probable such mechanism is indeed the predominantly centrifugally
driven MHD wind described above.  He also suggested that an additional
source of energy is required to produce jets that are sufficiently
powerful to be observed.  In this paper, we will provide qualitative
and quantitative evidence in support of this hypothesis.

It is now well established that weakly magnetized accretion disks are
subject to a magnetorotational instability (Velikhov\markcite{V59}
1959; Chandrasekhar\markcite{C60} 1960; Balbus \&
Hawley\markcite{BH91} 1991) of which the non-linear development is a
state of MHD turbulence (Brandenburg et~al.\markcite{BNST95} 1995;
Hawley, Gammie, \& Balbus\markcite{HGB95} 1995; Stone
et~al.\markcite{SHGB96} 1996).  For this reason, we include a
stability analysis as part of this investigation.  However, we do not
attempt to address the precise relation between MHD turbulence and MHD
winds.

The remainder of this paper is organized as follows.  In \S2, we
present the equations governing the equilibrium of the disk, and
describe the numerical solutions in terms of `standard', dimensionless
units.  We also discuss the stability of the equilibria.  In \S3, we
consider the dynamics of a transonic outflow in the corona of the
disk, and identify the potential difference that must be overcome by
such an outflow.  In \S4, we translate the results into physical
units.  Finally, in \S5, we discuss the implications of our analysis,
with an emphasis on cataclysmic variable (CV) systems.

\section{Equilibrium of the disk}

\subsection{Basic equations}

In this section we describe the local vertical structure of a
magnetized accretion disk such as might be found in a CV system.  The
disk is taken to be a steady, axisymmetric MHD flow in the
gravitational potential of a spherical mass $M$.  The usual
cylindrical polar coordinates $(r,\phi,z)$ are adopted.  In the limit
of a geometrically thin disk, the governing equations may be
simplified considerably by means of an asymptotic analysis
(Ogilvie\markcite{O97a} 1997a; hereafter, Paper~I) in which the small
parameter $\epsilon$ is a characteristic value of $H(r)/r$, where
$z=H(r)$ is the location of the upper surface of the disk at radius
$r$.  They then take the form of ordinary differential equations
(ODEs) in the vertical coordinate $z$ at each radius $r$ separately,
which describe the local vertical equilibrium of the disk.  There is
also an integral relation describing the global magnetic structure,
but here we are concerned only with the local aspects of the problem.

In the present work we do not repeat the formal asymptotic analysis,
but present the simplified, approximate equations that result from it.
The dynamical aspects of the equilibrium were considered in
Paper~I\markcite{O97a}, but the treatment of thermal and radiative
physics below is new.\footnote{The scalings correspond to the `weakly
magnetized disks' of Paper~I\markcite{O97a}, although equilibria
resembling the `strongly magnetized disks' will appear in the limit of
a strong magnetic field.}  We consider the equations below to afford a
minimal description of the problem while allowing quantitative
predictions to be made.

\subsubsection{Balance of forces}

The angular velocity of the fluid is written as
$\Omega=\Omega_0+\Omega_1$, where
\begin{equation}
\Omega_0=\left({{GM}\over{r^3}}\right)^{1/2}
\end{equation}
is the Keplerian value, which is independent of $z$, and $\Omega_1$ is
the deviation from Keplerian rotation, which depends on $z$, but is
much smaller than $\Omega_0$.  The radial component of the momentum
equation then becomes
\begin{equation}
-2\rho r\Omega_0\Omega_1={{B_z}\over{\mu_0}}{{\partial
B_r}\over{\partial z}}.
\end{equation}
(We write the permeability of free space as $\mu_0$, but will
substitute $\mu_0=4\pi$ later for calculations in CGS units.)  This
equation states that the radial tension force associated with the
bending magnetic field produces the deviation from Keplerian rotation.
Radial gradients of fluid and magnetic pressures are negligible, as
are the inertial effects of the accretion flow.  The vertical
component of the momentum equation is
\begin{equation}
0=-{{\partial p}\over{\partial z}}-{{B_r}\over{\mu_0}}{{\partial
B_r}\over{\partial z}}-\rho\Omega_0^2z.
\end{equation}
This equation states that both the pressure of the radial magnetic
field and the vertical gravitational force act to compress the disk.
The pressure of any toroidal magnetic field is neglected, as is the
self-gravitation of the disk.

\subsubsection{Magnetic field}

The solenoidal condition on the magnetic field implies
\begin{equation}
{{\partial B_z}\over{\partial z}}=0.
\end{equation}
The induction equation reduces to the condition of isorotation, ${\bf
B}\!\cdot\!\nabla\Omega=0$, which becomes
\begin{equation}
-{{3\Omega_0}\over{2r}}B_r+B_z{{\partial\Omega_1}\over{\partial
z}}=0.\label{eq4}
\end{equation}
The effects of any turbulence in the disk on the {\it mean} magnetic
field are neglected.  If the scale of the turbulent motions were small
compared to the scale on which the magnetic field varies, it might be
possible to describe the mean turbulent EMF in terms of an
$\alpha$-effect, representing the regeneration of the mean magnetic
field, and a $\beta$-effect, representing turbulent diffusion
(e.g.~Moffatt\markcite{M78} 1978).  However, there is no such
separation of scales in the present problem and the effects of any
turbulence are completely uncertain.  Although these effects may be
important in practice, our assumption of isorotation is the simplest
way to proceed in the face of this uncertainty.

\subsubsection{Thermal and radiative physics}

Instead of prescribing a polytropic relation between pressure and
density, as was done in Paper~I\markcite{O97a}, we now include thermal
and radiative physics.  The disk is assumed to be optically thick,
with Rosseland mean opacity $\kappa$.  The vertical radiative energy
flux is then
\begin{equation}
F=-{{16\sigma T^3}\over{3\kappa\rho}}{{\partial T}\over{\partial z}},
\end{equation}
and the energy equation may be written
\begin{equation}
{{\partial F}\over{\partial z}}={{3}\over{2}}\Omega_0\sigma_{r\phi},
\end{equation}
where $\sigma_{r\phi}$ is the component of the stress tensor
responsible for angular momentum transport and dissipation of energy.
(Note that the solutions we are considering are stable against
convection.)

For a complete specification of the problem, an equation of state must
be supplied, and also prescriptions for the opacity and stress.  We
use the ideal gas law,
\begin{equation}
p={{k\rho T}\over{\mu m_{\rm H}}},
\end{equation}
where $\mu$ is the mean molecular weight of the gas.  This assumes
that the radiation pressure is negligible.  For the stress, we adopt
the conventional $\alpha$-prescription,
\begin{equation}
\sigma_{r\phi}=\alpha p,\label{eq1}
\end{equation}
where $\alpha$ is a constant.  For the opacity, we assume a Kramers
law,
\begin{equation}
\kappa=\kappa_0\rho T^{-7/2},
\end{equation}
where $\kappa_0$ is a constant.  This is an adequate approximation
when free-free absorption is the dominant source of opacity, as is
usually the case in CV disks.  We consider more general opacity laws
in \S4 below.  While equation (\ref{eq1}) represents the most
uncertain element of this treatment, it does allow a comparison to be
made with previous work.

\subsubsection{Boundary conditions}

We solve the equations in $0<z<H$, and apply symmetry conditions
\begin{equation}
B_r=0
\end{equation}
and
\begin{equation}
F=0
\end{equation}
at $z=0$.  The magnetic field then has dipolar symmetry.  The boundary
conditions at $z=H$ are taken to be
\begin{equation}
\rho=0,
\end{equation}
\begin{equation}
T=0
\end{equation}
and
\begin{equation}
B_r=B_z\tan i,
\end{equation}
where $i$ is the angle of inclination of the magnetic field to the
vertical at the surface of the disk.

\subsubsection{Conservation of angular momentum}

A further relation, which allows quantities to be expressed in terms
of the accretion rate $\dot M$, is obtained from the conservation of
angular momentum.  Making the usual assumptions about the nature of
the boundary layer near the surface of the central object at radius
$r_*$ (e.g.~Pringle\markcite{P81} 1981), we obtain
\begin{equation}
r^2\Omega_0f\dot M=2\pi r^2\int\sigma_{r\phi}\,{\rm d}z,
\end{equation}
where
\begin{equation}
f=1-\left({{r_*}\over{r}}\right)^{1/2}.
\end{equation}
Then
\begin{equation}
F_{\rm s}={{3}\over{8\pi}}\Omega_0^2f\dot M,\label{eq8}
\end{equation}
where the subscript `s' denotes the value at the surface of the disk.

This is one equation in which a toroidal magnetic field might have a
significant effect.  If the disk has an outflow which exerts a torque
comparable to that due to $\sigma_{r\phi}$, it can be considered to
increase the effective value of $F_{\rm s}$ in this equation.
However, it is assumed that the rate at which mass is lost to the wind
is small compared to $\dot M$.  (See, e.g., Livio\markcite{L97} 1997
for observational evidence suggesting that this is indeed the case.)

\subsection{Transformation of the equations}

We now recast the equations in dimensionless form by means of the
following transformations.
\begin{eqnarray}
z&=&\tilde z\,H,\\
\Omega_1&=&\tilde\Omega_1\,\left({{H}\over{r}}\right)\Omega_0,\\
\rho&=&\tilde\rho\,\left({{3\alpha}\over{2}}\right)^{-1/3}H^{11/3}
\Omega_0^4\left({{\mu m_{\rm H}}\over{k}}\right)^{5/2}
\left({{16\sigma}\over{3\kappa_0}}\right)^{1/3},\\
p&=&\tilde
p\,\left({{3\alpha}\over{2}}\right)^{-1/3}H^{17/3}\Omega_0^6
\left({{\mu m_{\rm H}}\over{k}}\right)^{5/2}
\left({{16\sigma}\over{3\kappa_0}}\right)^{1/3},\\
{\bf B}&=&\tilde{\bf B}\,\left({{3\alpha}\over{2}}\right)^{-1/6}
\mu_0^{1/2}H^{17/6}\Omega_0^3
\left({{\mu m_{\rm H}}\over{k}}\right)^{5/4}
\left({{16\sigma}\over{3\kappa_0}}\right)^{1/6},\label{eq14}\\
T&=&\tilde T\,H^2\Omega_0^2\left({{\mu m_{\rm H}}\over{k}}\right),\\
F&=&\tilde
F\,\left({{3\alpha}\over{2}}\right)^{2/3}H^{20/3}\Omega_0^7
\left({{\mu m_{\rm H}}\over{k}}\right)^{5/2}
\left({{16\sigma}\over{3\kappa_0}}\right)^{1/3}.\label{eq9}
\end{eqnarray}
The dimensionless equations are
\begin{eqnarray}
{{\partial\tilde\Omega_1}\over{\partial\tilde z}}&=&
{{3\tilde B_r}\over{2\tilde B_z}},\label{eq2}\\
{{\partial\tilde p}\over{\partial\tilde z}}&=&-\tilde\rho\tilde z+
{{2\tilde\rho\tilde\Omega_1\tilde B_r}\over{\tilde B_z}},\\
{{\partial\tilde B_r}\over{\partial\tilde z}}&=&
-{{2\tilde\rho\tilde\Omega_1}\over{\tilde B_z}},\\
{{\partial\tilde T}\over{\partial\tilde z}}&=&
-\tilde\rho^2\tilde T^{-13/2}\tilde F,\\
{{\partial\tilde F}\over{\partial\tilde z}}&=&
\tilde p,\\
\tilde p&=&\tilde\rho\tilde T.\label{eq3}
\end{eqnarray}
We then have a fifth-order system of non-linear ODEs on $0<\tilde
z<1$, with boundary conditions
\begin{eqnarray}
\tilde B_r(0)&=&0,\\
\tilde F(0)&=&0,\\
\tilde\rho(1)&=&0,\\
\tilde T(1)&=&0,\\
\tilde B_r(1)&=&\tilde B_z\tan i,
\end{eqnarray}
and with two dimensionless parameters, $\tilde B_z$ and $i$, which may
be taken to be non-negative without loss of generality.  Note that the
dimensionless viscosity parameter $\alpha$ has been scaled out of the
equations.  We remark that a similar reduction of the equations could
be made for any equation of state, stress prescription and opacity law
that are `simple', in the sense of being monomials of the
thermodynamic variables.

Dimensionless variables written with tildes will be said to be
expressed in `standard' units.  These units allow the simplest and
most natural presentation of the equations and their solutions.  They
are based on the length scale $H$, the time scale $\Omega^{-1}$, and
mass and temperature scales derived from the coefficients appearing in
the equation of state and the opacity law; they also incorporate the
scalings of the solutions with $H/r$ and $\alpha$.  However, since $H$
is not known a priori, a further transformation is made in \S4 below
to express quantities in `physical' units based on the accretion rate.
For this reason, the interpretation of numerical results in the next
section should be regarded as provisional.  The final interpretation
is offered in \S4.

\subsection{Numerical solutions}

An expansion of the required solution of equations
(\ref{eq2})--(\ref{eq3}) in powers of $(1-\tilde z)$ is readily
obtained which allows the equations to be integrated smoothly out of
the singular point $\tilde z=1$ towards $\tilde z=0$.  Shooting is
then required in two dimensions to match the symmetry conditions on
$\tilde z=0$.  As discussed by Ogilvie\markcite{O97b} (1997b), for
certain values of the parameters there exist solutions in which the
magnetic field bends more than once as it passes through the disk.
Such `irregular' equilibria are known to be unstable and are not
discussed here.  The regular equilibria occupy a connected region in
the parameter space, as shown in Figure~1.  The edge of this region
has a form similar to that obtained for polytropic disks.  Equilibria
with any angle $i<30^\circ$ exist for any value of $\tilde B_z$, but
angles $i>30^\circ$ can be achieved only if the magnetic field is
sufficiently strong.  As the edge of the solution manifold is
approached, the effective gravitational acceleration parallel to the
magnetic field at the surface of the disk tends to zero, which in a
time-dependent situation would lead to the dynamical disruption of the
disk.

\placefigure{fig1}

The equilibria exhibit a wide variety of behavior in different parts
of the parameter space, and certain features of this should be
explained.  First, when $i=0$, the equilibria are unaffected by the
magnetic field and all take the form of the unmagnetized solution.
Secondly, when $i$ is fixed, with $0<i<30^\circ$, and the limit
$\tilde B_z\to0$ is taken, the equilibria eventually become physically
unrealistic.  Much of the disk becomes almost evacuated, with mass
concentrated near the equatorial plane and near the surface.  This
behavior is the only way to prevent the magnetic field lines from
bending many times as they pass through the disk, when the field is
very weak.  While these equilibria are almost certainly unstable to
overturning, this behavior occurs only when they are already unstable
to the magnetorotational instability, as described in \S2.4 below.
Thirdly, when $i$ is fixed, with $i>0$, and the limit $\tilde
B_z\to\infty$ is taken, the equilibria approach a strongly magnetized
limit in which they are compressed only by the Lorentz force and not
by gravity.  We find the asymptotic form of the solution to be
\begin{eqnarray}
\tilde\Omega_1(\tilde z)&=&
-\tilde B_z^{12/19}(\tan i)^{-7/19}\,\lambda+O(1),\\
\tilde\rho(\tilde z)&\sim&(\tilde B_z\tan i)^{26/19}\,y_1(\tilde z),\\
\tilde p(\tilde z)&\sim&(\tilde B_z\tan i)^2\,y_2(\tilde z),\\
\tilde B_r(\tilde z)&\sim&(\tilde B_z\tan i)\,y_3(\tilde z),\\
\tilde T(\tilde z)&\sim&(\tilde B_z\tan i)^{12/19}\,y_4(\tilde z),\\
\tilde F(\tilde z)&\sim&(\tilde B_z\tan i)^2\,y_5(\tilde z),
\end{eqnarray}
where $\lambda$ is an eigenvalue of the non-linear system
\begin{eqnarray}
y_2^\prime&=&-2\lambda y_1y_3,\\
y_3^\prime&=&2\lambda y_1,\\
y_4^\prime&=&-y_1^2y_4^{-13/2}y_5,\\
y_5^\prime&=&y_2,\\
y_2&=&y_1y_4,
\end{eqnarray}
with boundary conditions
\begin{eqnarray}
y_3(0)&=&0,\\
y_5(0)&=&0,\\
y_1(1)&=&0,\\
y_4(1)&=&0,\\
y_3(1)&=&1.
\end{eqnarray}
Here, a prime denotes differentiation with respect to the argument.
The numerically determined solution is $\lambda\approx1.916$.  Also
required is the quantity $y_5(1)\approx0.1607$.  This limiting
behavior of the equilibria is used in \S4 below.

The plasma beta, being the ratio of the gas pressure to the magnetic
pressure, is often taken to be a dimensionless, inverse measure of the
strength of the magnetic field.  However, as explained in
Paper~I\markcite{O97a}, this is not appropriate in the case of
strongly magnetized accretion disks.  In the limit $\tilde
B_z\to\infty$, the plasma beta, evaluated on the equatorial plane,
tends to a constant value which depends on $i$.  This is because the
nature of the vertical equilibrium in the strongly magnetized limit
requires that the gas pressure increase in proportion to the magnetic
pressure.

\subsection{Magnetorotational stability of the equilibria}

If the magnetic field is sufficiently weak, the equilibria are
expected to be unstable to the magnetorotational instability (Balbus
\& Hawley\markcite{BH91} 1991).  The instability of a thin disk
containing a bending poloidal magnetic field has been analyzed by
Ogilvie\markcite{O97b} (1997b), who found that the curve of marginal
stability in the parameter space could be located by solving the
equations for an equilibrium possessing a mode with zero frequency and
zero radial wavenumber.  In terms of the Lagrangian displacement
${\bf\xi}$, these are
\begin{equation}
3\Omega_0^2\rho\,\xi_r=-B_z{{\partial}\over{\partial z}}
\left(B_z{{\partial\xi_r}\over{\partial z}}-
B_r{{\partial\xi_z}\over{\partial z}}\right)
\end{equation}
and
\begin{equation}
\Omega_0^2z{{\partial\rho}\over{\partial z}}\,\xi_z=
{{\partial\,\delta\Pi}\over{\partial z}}
-\rho\Omega_0^2z{{\partial\xi_z}\over{\partial z}},
\end{equation}
where
\begin{equation}
\delta\Pi=\rho\Omega_0^2z\,\xi_z+
B_rB_z{{\partial\xi_r}\over{\partial z}}-
(\gamma p+B_r^2){{\partial\xi_z}\over{\partial z}}
\end{equation}
is the Eulerian perturbation of total pressure.  This assumes that the
perturbations are adiabatic, with adiabatic exponent $\gamma$.  The
relevant mode is the first mode of odd symmetry, which satisfies the
boundary conditions
\begin{equation}
\xi_r={{\partial\xi_z}\over{\partial z}}=0
\end{equation}
at $z=0$, and
\begin{equation}
{{\partial\xi_r}\over{\partial z}}-
\tan i\,{{\partial\xi_z}\over{\partial z}}=0
\end{equation}
at $z=H$.  The dimensionless form of these equations is identical
except for the inclusion of the tildes and the omission of $\Omega_0$.

The curve of marginal stability, for the case $\gamma=5/3$, is shown
in Figure~2.  Again, this is qualitatively similar to curves obtained
for polytropic disks.

\placefigure{fig2}

As discussed by Ogilvie\markcite{O97b} (1997b), it is expected that
the equilibria that are stable to the magnetorotational instability
are also stable to the magnetoconvective (Parker), interchange, and
bending instabilities, unless the magnetic field provides most of the
support against the radial gravitational force.  Although a weak,
global, non-axisymmetric instability may remain, this is unlikely to
be dynamically important in a thin disk.

\section{Dynamics of an outflow in the corona of the disk}

Following Paper~I\markcite{O97a}, we now analyze the dynamics of a
transonic outflow in the region immediately above the disk, which we
refer to as the `corona', and which is defined by $H<z\ll r$.  The
density of the wind is very much smaller than that of the disk,
possibly by a factor $O(\epsilon^4)$, and to match the solutions in
detail would require high-order asymptotics well beyond the scope of
this analysis.  Instead, we consider that the disk acts as a reservoir
which can supply any reasonable mass flux to the wind, the
corresponding velocity field required in the disk being extremely
small.

In the corona, the magnetic field is force-free to a very good
approximation, and the field lines are straight on the length scale
$H$.  They act as rigid channels for the wind and also enforce
isorotation.  The dynamics of the outflow in this region depends
critically on the centrifugal--gravitational potential $\Phi^{\rm
cg}$, which may be computed as follows.  Since the magnetic field is
force-free, $\partial B_r/\partial z=0$, so that $B_r=B_z\tan i$
throughout the corona.  Equation (\ref{eq4}) may then be integrated to
give
\begin{equation}
\Omega_1=\Omega_{1{\rm s}}+{{3\Omega_0(z-H)\tan i}\over{2r}},
\end{equation}
where, again, the subscript `s' denotes the value at the surface of
the disk.  The effective gravitational acceleration is
\begin{equation}
{\bf g}=2r\Omega_0\Omega_1\,{\bf e}_r-\Omega_0^2z\,{\bf e}_z,
\end{equation}
and its component measured parallel to the magnetic field (and towards
the surface of the disk) is
\begin{equation}
g_\|=-(3\tan^2i-1)\Omega_0^2(z-H)\cos i+
\Omega_0(\Omega_0H-2\Omega_{1{\rm s}}r\tan i)\cos i.
\end{equation}
If, as we assume, $i>30^\circ$, then $g_\|$ decreases linearly with
increasing $z$ and goes to zero at $z=z_{\rm sonic}$, given by
\begin{equation}
z_{\rm sonic}=H+{{(\Omega_0H-2\Omega_{1{\rm s}}r\tan i)}\over
{(3\tan^2i-1)\Omega_0}}.\label{eq5}
\end{equation}
This name is appropriate because, as shown below, $z=z_{\rm sonic}$ is
the expected location of the sonic point of a transonic wind.
Consider a single magnetic field line, and let $z$ parametrize the
position along it.  When $g_\|$ is integrated along the field line, it
yields the centrifugal--gravitational potential
\begin{equation}
\Phi^{\rm cg}=-{\textstyle{{1}\over{2}}}(3\tan^2i-1)\Omega_0^2
(z-z_{\rm sonic})^2,
\end{equation}
as measured from the sonic point.  In particular, the potential
difference between the sonic point and the surface of the disk is
\begin{equation}
\Delta\Phi={{(\Omega_0H-2\Omega_{1{\rm s}}r\tan i)^2}\over
{2(3\tan^2i-1)}}.
\end{equation}
When $i$ approaches $30^\circ$ (from above), the height of the sonic
point increases without bound, as does the potential difference.
However, these expressions are valid only if the height of the sonic
point calculated from equation (\ref{eq5}) satisfies $z_{\rm sonic}\ll
r$, since otherwise the quadratic approximation to the potential
ceases to be valid and the curvature of the magnetic field lines
should also be taken into account.  Within this approximation, the
potential difference is always small compared to $GM/r$.  (By
comparison, when $i<30^\circ$, the potential difference is comparable
to $GM/r$.)

We now consider the dynamics of a transonic wind flowing along the
magnetic field lines.  The wind is treated as isothermal, since the
optical depth is presumably small in the corona, and therefore
\begin{equation}
p=c^2\rho,
\end{equation}
where
\begin{equation}
c=\left({{kT}\over{\mu m_{\rm H}}}\right)^{1/2}
\end{equation}
is the isothermal sound speed.  Mass conservation requires that the
mass flux density
\begin{equation}
\rho u={\cal F}={\rm constant}
\end{equation}
be constant along the flow.  The Bernoulli equation states that
\begin{equation}
{\textstyle{{1}\over{2}}}u^2+c^2\ln\rho+\Phi^{\rm cg}={\rm constant}
\end{equation}
is also constant following the flow.  On differentiating this equation
we obtain
\begin{equation}
(u^2-c^2){{{\rm d}\ln u}\over{{\rm d}z}}=
(3\tan^2i-1)\Omega_0^2(z-z_{\rm sonic}),
\end{equation}
which demonstrates that the sonic transition must occur at $z=z_{\rm
sonic}$.  In terms of the Mach number ${\cal M}=u/c$, we then have
\begin{equation}
{\textstyle{{1}\over{2}}}({\cal M}^2-1)-\ln{\cal M}=-\Phi^{\rm cg}/c^2,
\end{equation}
and, in particular,
\begin{equation}
{\textstyle{{1}\over{2}}}({\cal M}_{\rm s}^2-1)-\ln{\cal M}_{\rm s}=
\Delta\Phi/c^2\label{eq6}
\end{equation}
is a transcendental equation for the Mach number ${\cal M}_{\rm s}$ at
the surface of the disk.  The mass flux density is then
\begin{equation}
{\cal F}={\cal M}_{\rm s}\rho_{\rm s}c,
\end{equation}
where $\rho_{\rm s}$ is the density of the wind at the surface of the
disk.  When $\Delta\Phi\gg c^2$, an approximate solution is
\begin{equation}{\cal M}_{\rm s}\approx
\exp(-\Delta\Phi/c^2-{\textstyle{{1}\over{2}}}),\label{eq15}
\end{equation}
which implies
\begin{equation}
{\cal F}\approx\rho_{\rm s}c\,
\exp(-\Delta\Phi/c^2-{\textstyle{{1}\over{2}}}).\label{eq7}
\end{equation}
The numerical solution of equation (\ref{eq6}) in Figure~3 shows that
this approximation is accurate even for moderate values of
$\Delta\Phi/c^2$.

\placefigure{fig3}

Evidently the meaning of equation (\ref{eq7}) is that the outflow is
severely choked if the potential difference is much larger than $c^2$.
For any equilibrium we can compute $\Delta\Phi$ and deduce the sound
speed -- and therefore the temperature -- required in the corona for
the outflow not to be suppressed.  The height of the sonic point and
the potential difference are plotted as functions of $i$ for various
values of $\tilde B_z$ in Figure~4.  These equilibria are all stable
to the magnetorotational instability.  As $i$ increases from
$30^\circ$, the sonic point approaches the surface of the disk and the
potential difference decreases, until the equilibrium ceases to exist.
When the magnetic field is stronger, the sonic point is more distant
and the potential difference is larger.  However, for any field
strength, an angle $i$ can be found for which the potential difference
is arbitrarily small.  In the next section we show that this property
is lost when the solutions are expressed in more physical units which
take account of the compression of the disk by the magnetic field.

\placefigure{fig4}

\section{The solutions expressed in physical units}

\subsection{Definition of units}

Although the dimensionless forms defined in \S2.2 are convenient for
the purposes of computation, one would like to express the magnetic
field, for example, in units of gauss rather than in the dimensionless
form $\tilde{\bf B}$.  The relation between ${\bf B}$ and $\tilde{\bf
B}$ involves $H$ which is not known a priori.  However, the accretion
rate $\dot M$ is determined a priori in the sense of being of a global
property of the disk, independent of radius, whose value can be
estimated observationally.  From the numerical solutions, we know
$\tilde F_{\rm s}$ for each equilibrium, which allows us to write
\begin{equation}
{{3}\over{8\pi}}\Omega_0^2f\dot M=
\tilde F_{\rm s}\,\left({{3\alpha}\over{2}}\right)^{2/3}H^{20/3}
\Omega_0^7\left({{\mu m_{\rm H}}\over{k}}\right)^{5/2}
\left({{16\sigma}\over{3\kappa_0}}\right)^{1/3},
\end{equation}
combining equations (\ref{eq8}) and (\ref{eq9}).  We then have
\begin{equation}
H=\tilde F_{\rm s}^{-3/20}\,U_H,
\end{equation}
where
\begin{equation}
U_H=\left({{3}\over{8\pi}}\right)^{3/20}
\left({{3\alpha}\over{2}}\right)^{-1/10}\Omega_0^{-3/4}
\left({{\mu m_{\rm H}}\over{k}}\right)^{-3/8}
\left({{16\sigma}\over{3\kappa_0}}\right)^{-1/20}f^{3/20}\dot M^{3/20}
\end{equation}
is a suitable unit of length.  Similarly, we may write
\begin{equation}
{\bf B}=\tilde{\bf B}\tilde F_{\rm s}^{-17/40}\,U_B
\end{equation}
and
\begin{equation}
T=\tilde T\tilde F_{\rm s}^{-3/10}\,U_T,
\end{equation}
where
\begin{equation}
U_B=\left({{3}\over{8\pi}}\right)^{17/40}
\left({{3\alpha}\over{2}}\right)^{-9/20}\mu_0^{1/2}\Omega_0^{7/8}
\left({{\mu m_{\rm H}}\over{k}}\right)^{3/16}
\left({{16\sigma}\over{3\kappa_0}}\right)^{1/40}
f^{17/40}\dot M^{17/40}\label{eq16}
\end{equation}
and
\begin{equation}
U_T=\left({{3}\over{8\pi}}\right)^{3/10}
\left({{3\alpha}\over{2}}\right)^{-1/5}\Omega_0^{1/2}
\left({{\mu m_{\rm H}}\over{k}}\right)^{1/4}
\left({{16\sigma}\over{3\kappa_0}}\right)^{-1/10}
f^{3/10}\dot M^{3/10}
\end{equation}
are suitable units of magnetic field strength and of temperature.

In terms of CGS units, we find
\begin{eqnarray}
U_H&\approx&1.2\times10^8\,\alpha^{-1/10}M_1^{-3/8}R_{10}^{9/8}
f^{3/20}\dot M_{16}^{3/20}\,{\rm cm},\\
U_B&\approx&1.0\times10^3\,\alpha^{-9/20}M_1^{7/16}R_{10}^{-21/16}
f^{17/40}\dot M_{16}^{17/40}\,{\rm gauss},\\
U_T&\approx&1.3\times10^4\,\alpha^{-1/5}M_1^{1/4}R_{10}^{-3/4}
f^{3/10}\dot M_{16}^{3/10}\,{\rm K},
\end{eqnarray}
where $M_1=M/M_\odot$, $R_{10}=R/(10^{10}\,{\rm cm})$, $\dot
M_{16}=\dot M/(10^{16}\,{\rm g}\,{\rm s}^{-1})$, and we have used the
values $\mu\approx0.6$ and $\kappa_0\approx6.4\times10^{22}\,{\rm
cm}^2\,{\rm g}^{-1}$ (cf.~Novikov \& Thorne\markcite{NT73} 1973)
appropriate for CV disks.

\subsection{Solutions with a purely vertical magnetic field}

When the magnetic field is purely vertical, it does not affect the
equilibrium.  The solution is the same for all values of $\tilde B_z$,
and is found numerically to have $\tilde F_{\rm s}\approx0.0007041$.
We then obtain
\begin{equation}
H\approx2.971\,U_H\approx3.5\times10^8\,\alpha^{-1/10}M_1^{-3/8}
R_{10}^{9/8}f^{3/20}\dot M_{16}^{3/20}\,{\rm cm}.
\end{equation}
We note that this is larger by a factor of approximately 2 than the
value quoted by, for example, Frank, King, \& Raine\markcite{FKR85}
(1985), which was based on order-of-magnitude estimates (although we
define $H$ to be the true semi-thickness rather than an approximate
scale height).  The magnetic field strength for marginal
magnetorotational stability is found to be $\tilde B_z\approx0.06327$,
or
\begin{equation}
B_z\approx1.383\,U_B\approx1.4\times10^3\,\alpha^{-9/20}
M_1^{7/16}R_{10}^{-21/16}f^{17/40}\dot M_{16}^{17/40}\,{\rm
gauss}.\label{eq13}
\end{equation}
If the magnetic field results from the aligned dipole field of a
central white dwarf of radius $5.0\times10^8\,{\rm cm}$, a value of
$1.4\times10^3\,{\rm gauss}$ at a radius of $10^{10}\,{\rm cm}$ in the
disk would correspond to a value of $2.2\times10^7\,{\rm gauss}$ at
the poles of the white dwarf.  This means that the disk would be
unstable at this radius in many cases of astrophysical interest, but
may be stable in some systems such as V1500 Cyg.

\subsection{Solutions with an inclined magnetic field}

We now redraw Figure~2 in physical units, using $\tilde B_z\tilde
F_{\rm s}^{-17/40}$ as the ordinate rather than $\tilde B_z$.
Unfortunately the resulting graph (Figure~5), while more physically
meaningful, is more difficult to interpret.  The mapping from $\tilde
B_z$ to $\tilde B_z\tilde F_{\rm s}^{-17/40}$ is not one-to-one, with
the result that the solution manifold folds over on itself on the
left-hand side of Figure~5.  However, the `folded' solutions are
unstable and this detail need not be pursued here.  More interesting
is the way in which the curve on which the equilibria cease to exist
is transformed relative to Figure~2.  This distortion occurs because
equilibria with highly inclined magnetic field lines are strongly
compressed by the Lorentz force and therefore have a higher pressure
than the unmagnetized solution with the same $H$.  The stress and
torque are correspondingly increased, so that $\tilde F_{\rm s}$ is
large, and therefore $B_z/U_B$ is smaller than might be expected.

\placefigure{fig5}

The existence of the fold means that, when the magnetic field strength
is reduced below the stability boundary, the equilibria eventually
cease to exist.  However, it is of little importance whether, in this
part of the parameter space, no stable solution exists or no solution
whatever exists.  The result is likely to be a state of MHD
turbulence.

The height of the sonic point and the potential difference, expressed
in physical units, are plotted as functions of $i$ for various values
of $B_z/U_B$ in Figure~6.  The height of the sonic point is compared
with $H$, both expressed in units of $U_H$.  For the potential
difference we introduce a physical unit
\begin{equation}
U_\Phi={{kU_T}\over{\mu m_{\rm H}}}\approx1.8\times10^{12}\,
\alpha^{-1/5}M_1^{1/4}R_{10}^{-3/4}f^{3/10}\dot M_{16}^{3/10}\,
{\rm cm}^2\,{\rm s}^{-2}.
\end{equation}
Then $\Delta\Phi/U_\Phi$ can be compared with $T_{\rm c}/U_T$, where
$T_{\rm c}$ is the central temperature of the disk.

\placefigure{fig6}

In the range $1.55\la B_z/U_B\la1.93$, there exist stable equilibria
such that, when $i$ is increased from $30^\circ$, the potential
difference falls rapidly from infinity to zero as the sonic point
approaches the surface of the disk, and then the equilibria cease to
exist.  This behavior is similar to the interpretation offered
earlier, based on Figure~4 and standard units.  There are also
equilibria with more highly inclined magnetic field lines.  However,
for $B_z/U_B\ga1.93$, a different behavior is found.  As $i$ is
increased from $30^\circ$, the equilibria continue to exist for all
angles up to $90^\circ$.  The potential difference has a minimum,
typically in the range $38^\circ\la i\la43^\circ$, and then increases
again.  The sonic point at first approaches the surface of the disk,
but then recedes from it.  Moreover, as $B_z/U_B$ is increased, the
minimum potential difference increases very rapidly and quickly
exceeds a value corresponding to the central temperature of the disk.

\subsection{Limiting behavior for strongly magnetized disks}

When $B_z/U_B$ is sufficiently large, the limiting form of the
equilibria given in \S2.3 applies.  We then find that the potential
difference is
\begin{equation}
{{\Delta\Phi}\over{U_\Phi}}\sim C_1
\left({{B_z}\over{U_B}}\right)^{84/19}
{{(\tan i)^{84/19}}\over{(3\tan^2i-1)}},\label{eq12}
\end{equation}
where
\begin{equation}
C_1=2\lambda^2[y_5(1)]^{30/19}\approx0.4095.
\end{equation}
This increases very rapidly with increasing magnetic field strength.
The minimum with respect to $i$ occurs at
\begin{equation}
i=\arctan\sqrt{{14}\over{23}}\approx37.96^\circ.
\end{equation}
Similarly, the height of the sonic point is
\begin{equation}
{{z_{\rm sonic}}\over{U_H}}\sim C_2
\left({{B_z}\over{U_B}}\right)^{42/19}
{{(\tan i)^{42/19}}\over{(3\tan^2i-1)}},
\end{equation}
where
\begin{equation}
C_2=2\lambda[y_5(1)]^{15/19}\approx0.9050,
\end{equation}
and this has a minimum with respect to $i$ at
\begin{equation}
i=\arctan\sqrt{{7}\over{2}}\approx74.05^\circ.
\end{equation}
To show that these power laws and characteristic angles are relatively
insensitive to the opacity law, the analysis can be repeated using a
more general relation of the form
\begin{equation}
\kappa=\kappa_0\rho^xT^y.
\end{equation}
We then find that the potential difference is
\begin{equation}
{{\Delta\Phi}\over{U_\Phi}}\sim C_1
\left({{B_z}\over{U_B}}\right)^{4(5+2x-y)/(5+x-y)}
{{(\tan i)^{4(5+2x-y)/(5+x-y)}}\over{(3\tan^2i-1)}},
\end{equation}
where $C_1$ depends on $x$ and $y$, and of course the units $U_\Phi$,
$U_B$, etc., are differently but analogously defined.  This shows that
the potential difference typically scales as approximately the fourth
power of the magnetic field strength.  The minimum with respect to $i$
occurs at
\begin{equation}
i=\arctan\sqrt{{2(5+2x-y)}\over{3(5+3x-y)}},\label{eq17}
\end{equation}
provided that
\begin{equation}
{{5+3x-y}\over{5+x-y}}>0.\label{eq10}
\end{equation}
Similarly, the height of the sonic point is
\begin{equation}
{{z_{\rm sonic}}\over{U_H}}\sim C_2
\left({{B_z}\over{U_B}}\right)^{2(5+2x-y)/(5+x-y)}
{{(\tan i)^{2(5+2x-y)/(5+x-y)}}\over{(3\tan^2i-1)}},
\end{equation}
and this has a minimum with respect to $i$ at
\begin{equation}
i=\arctan\sqrt{{5+2x-y}\over{3x}},
\end{equation}
provided that
\begin{equation}
{{x}\over{5+x-y}}>0.\label{eq11}
\end{equation}
Criterion (\ref{eq10}) is satisfied for most reasonable opacity laws,
as shown in Figure~7.  In the case of electron-scattering opacity, for
which $x=y=0$, the potential difference scales as $(B_z/U_B)^4$ and is
minimized with respect to $i$ at $i\approx39.23^\circ$.  Criterion
(\ref{eq11}) marginally fails to be satisfied, however, so that the
height of the sonic point continues to decrease as $i$ approaches
$90^\circ$.

\placefigure{fig7}

In Figure~8 we compare the true minimum of the potential difference
with respect to $i$ with the value predicted from equation
(\ref{eq12}).  This shows that the limiting behavior described in this
section is achieved very rapidly as the vertical magnetic field
strength is increased beyond the stability boundary.

\placefigure{fig8}

\section{Discussion}

In this paper we have solved for the local vertical structure of a
magnetized accretion disk such as might be found in a CV system.  The
magnetic field was assumed to enforce isorotation, and the deviation
from Keplerian rotation was taken fully into account.  Angular
momentum transport in the disk was parametrized by an
$\alpha$-prescription in the conventional way, and a Kramers opacity
law was assumed to hold.  We have shown that, when quantities are
expressed in physical units, equilibria that are magnetorotationally
stable can be found in which the angle of inclination, $i$, of the
magnetic field to the vertical at the surface of the disk has any
value in the range $0\le i<90^\circ$ if the magnetic field is
sufficiently strong.

We have analyzed the dynamics of a transonic outflow in the corona of
the disk when $i>30^\circ$, and, in particular, have shown that a
certain potential difference must be overcome by such an outflow.
When the equilibria are very close to the magnetorotational stability
boundary, the potential difference is relatively small and can in fact
be made arbitrarily small by approaching the edge of the solution
manifold.  For more strongly magnetized disks, however, {\it the
potential difference increases faster than the fourth power of the
magnetic field strength, and is minimized with respect to $i$ at
$i\approx38^\circ$}.  These properties are relatively insensitive to
the opacity law.

We have used a local representation which neglects the global form of
the disk and wind, focusing instead on the vertical structure at a
single radius.  In this respect our analysis is comparable with that
of Wardle \& K\"onigl\markcite{WK93} (1993).  However, in the limit in
which ambipolar diffusion is negligible, as is the case in the
accretion disks of many classes of objects such as CVs, no meaningful
solution of their equations can be obtained, even though some of their
equations correspond in this limit to the ones we have solved.  To
insist that the disk achieves a balance between the outward diffusion
of magnetic flux and its inward advection by the accretion flow is
certainly attractive, but constrains the problem in such a way that it
appears that diffusion is driving the outflow.

Certain other features of our analysis should be emphasized.  First,
we do {\it not} assume that the wind accounts for all, or even most,
of the angular momentum transport in the disk.  In the case of CV
disks, for example, we can be confident that the agent supplying most
of the torque on the disk is not an MHD wind; see the discussion by
Livio\markcite{L97} (1997).  Secondly, we consider the disk to have a
well-defined surface which is hardly affected by the presence or
absence of an outflow above it.  We do not attempt to make a smooth
transition between the disk and the wind, except in requiring the
continuity of the mass and magnetic fluxes, because the density of the
wind is so small that a detailed matching procedure would require
high-order asymptotics.  Thirdly, we do not consider ambipolar
diffusion, Ohmic resistivity or turbulent diffusion of the magnetic
field.  Instead, we assume that isorotation holds on magnetic field
lines and we allow for a slow accretion of magnetic flux by the disk.
Finally, we have considered an optically thick disk rather than
assuming the disk to be isothermal.  We note that, if the disk is
isothermal, it is easier to obtain an outflow of reasonable strength
without postulating a hot corona, because the region above the disk
always has the same temperature as the central temperature.

The interpretation of our results depends to some extent on the class
of accretion disks under consideration, and on the source of the mean
magnetic field.  In the case of CV disks, for example, the magnetic
field of the white dwarf can be important and may be sufficient to
make at least the inner part of the disk magnetorotationally stable
(cf.~eq. [\ref{eq13}]).  In that case, even if the magnetic field
lines are inclined such that $i>30^\circ$, our results suggest that
the potential difference may be too large to allow a significant
outflow unless the disk has a hot corona or access to an additional
source of energy, such as nuclear burning, in accordance with the
hypothesis of Livio\markcite{L97} (1997).  However, the interaction
between a magnetized central object and the disk may be more
complicated than we have allowed for in this paper (cf.~Livio \&
Pringle\markcite{LP92} 1992; Miller \& Stone\markcite{MS97} 1997).  In
other cases, magnetic flux may be accreted from the environment or may
be a remnant of the formation of the disk.  In disks that are
magnetorotationally unstable, the dynamics of the mean magnetic field
is much less certain, but it may be possible for these systems to
regulate the distribution of magnetic flux so as to remain close to
the stability limit, and thereby avoid incurring a very large
potential difference.

\acknowledgments

GIO thanks the Space Telescope Science Institute for its hospitality.
ML acknowledges support from NASA Grant NAGW-2678.

\newpage

\figcaption[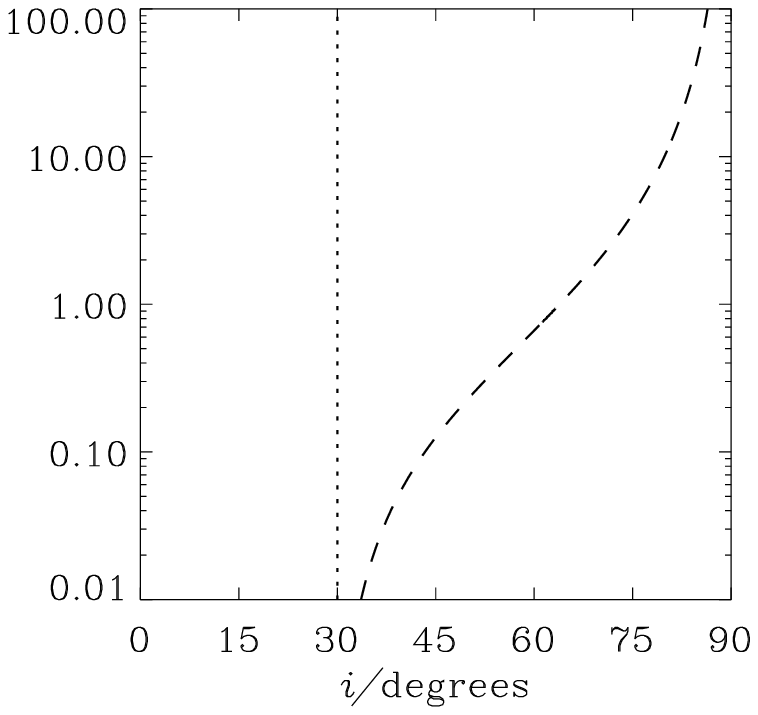]{Parameter space of equilibrium solutions.
$\tilde B_z$ is the vertical magnetic field in the disk, expressed in
standard units (eq. [\ref{eq14}]), and $i$ is the angle of inclination
of the magnetic field to the vertical at the surface of the disk.
{\it Dashed line\/}: edge of the solution manifold; regular equilibria
exist everywhere to the left of this line.  {\it Dotted line\/}: drawn
at an angle $i=30^\circ$, which is critical to the analysis in \S3.
(In this graph only, a logarithmic scale of magnetic field strength is
used in order to give a broad perspective on the parameter
space.)\label{fig1}}

\figcaption[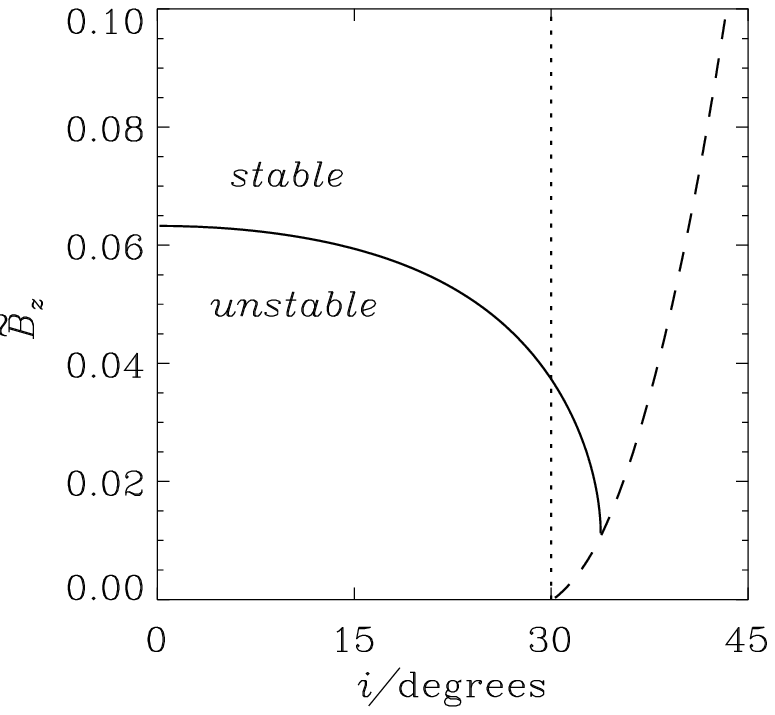]{{\it Solid line\/}: stability boundary to the
magnetorotational instability, for the case $\gamma=5/3$.  {\it Dashed
line\/}: edge of the solution manifold, as in Figure~1.  (Note that
the axes are quite different from those of Figure~1.)\label{fig2}}

\figcaption[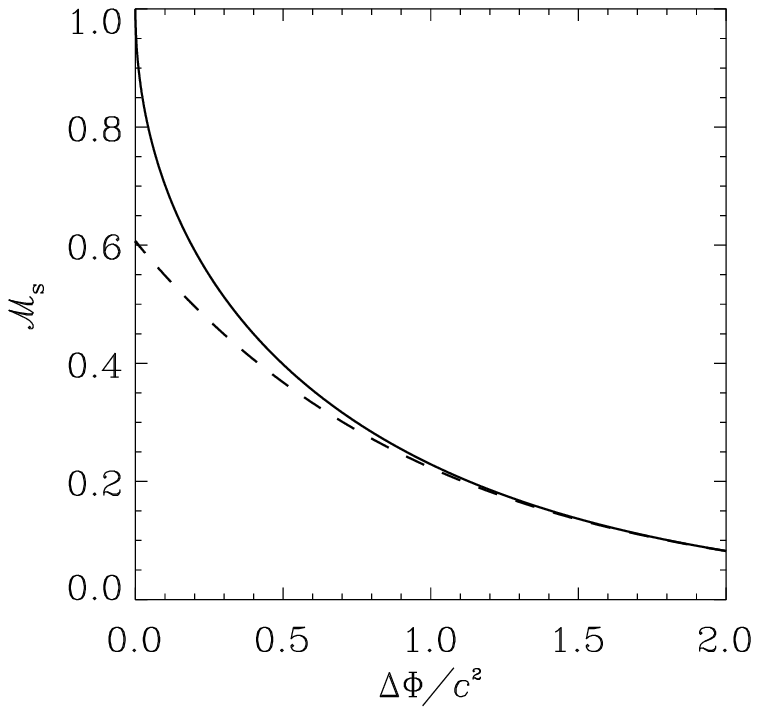]{Dependence of the Mach number ${\cal M}_{\rm s}$
of the wind at the surface of the disk on the potential difference
$\Delta\Phi$ between the sonic point and the surface, expressed in
units of $c^2$, where $c$ is the isothermal sound speed.  The solid
line denotes the true solution, while the dashed line represents the
approximation given in equation (\ref{eq15}).\label{fig3}}

\figcaption[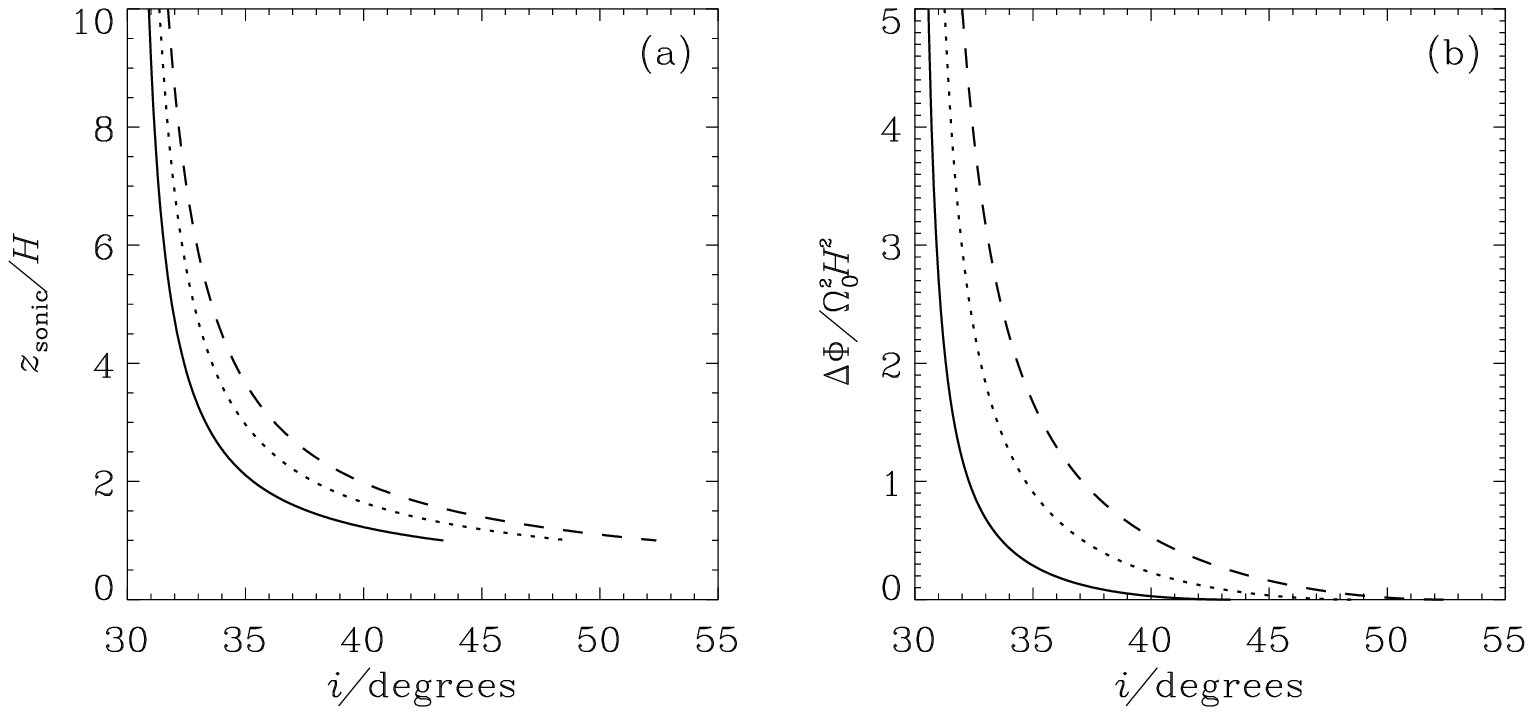]{Variation with $i$ of the height of the sonic
point ({\it a\/}) and the potential difference between the sonic point
and the surface ({\it b\/}), both expressed in standard units, for
equilibria with $\tilde B_z=0.1$ ({\it solid line\/}), $0.2$ ({\it
dotted line\/}) and $0.3$ ({\it dashed line\/}).\label{fig4}}

\figcaption[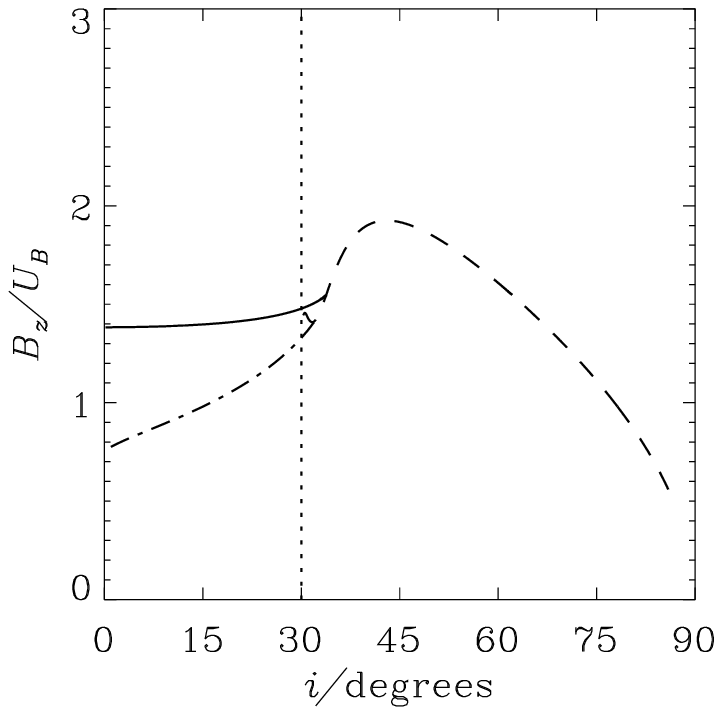]{Parameter space with the vertical magnetic field
$B_z$ expressed in physical units (eq. [\ref{eq16}]).  {\it Solid
line\/}: stability boundary.  {\it Dashed line\/}: edge of the
solution manifold.  {\it Dot-dashed line\/}: fold in the solution
manifold.  No regular equilibria are found below a curve which
consists of the dot-dashed line continued by the dashed
line.\label{fig5}}

\figcaption[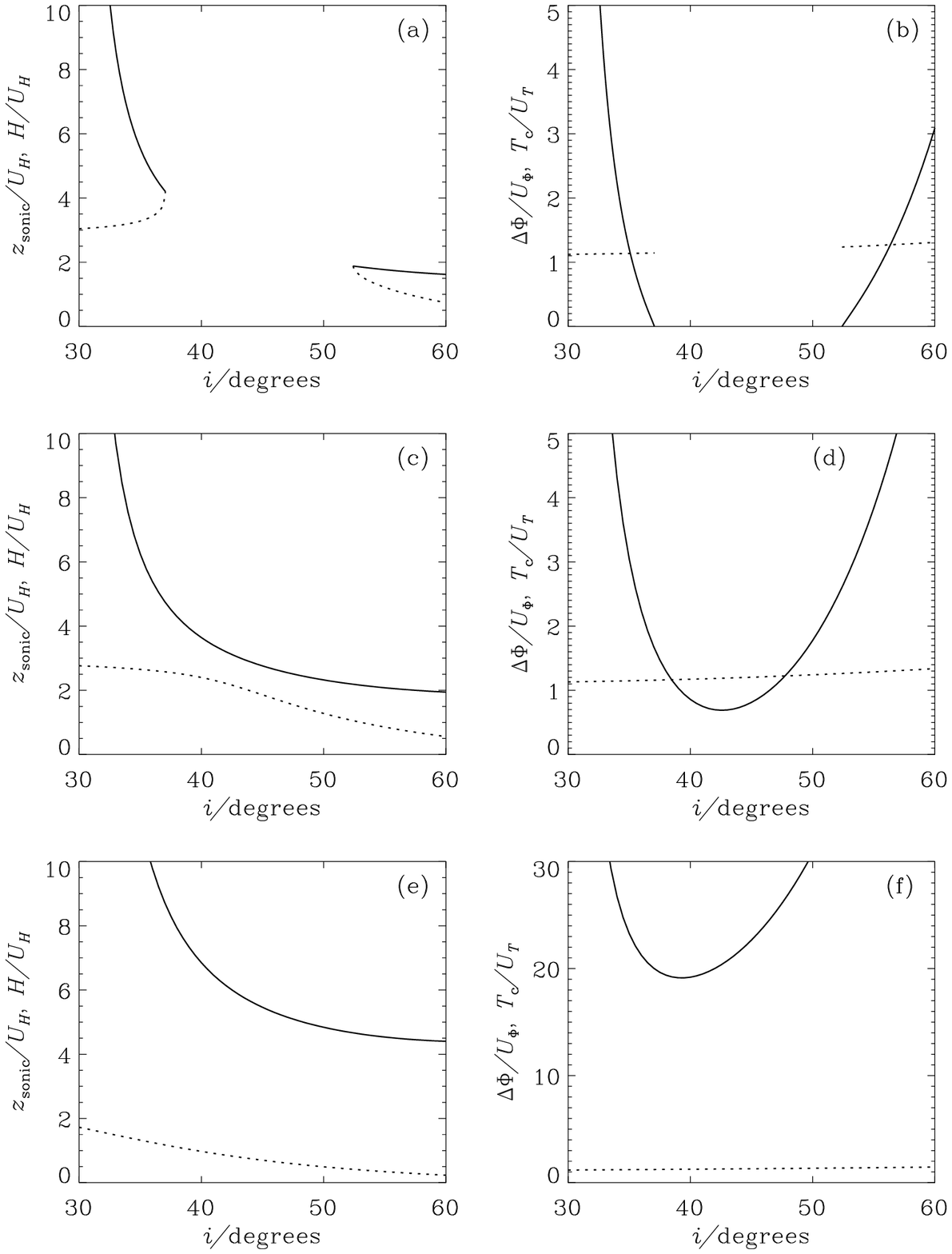]{{\it Panels a, c, and e\/}: height of the sonic
point ({\it solid line\/}) compared with $H$ ({\it dotted line\/}),
both expressed in physical units, as functions of $i$, for equilibria
with $B_z/U_B=1.8$ ({\it a\/}), $2.0$ ({\it c\/}), and $3.0$ ({\it
e\/}).  {\it Panels b, d, and f\/}: potential difference ({\it solid
line\/}) compared with the central temperature ({\it dotted line\/}),
both expressed in physical units, for equilibria with $B_z/U_B=1.8$
({\it b\/}), $2.0$ ({\it d\/}), and $3.0$ ({\it f\/}).  Note the
different scale for the potential difference and temperature in panel
{\it f\/}.\label{fig6}}

\figcaption[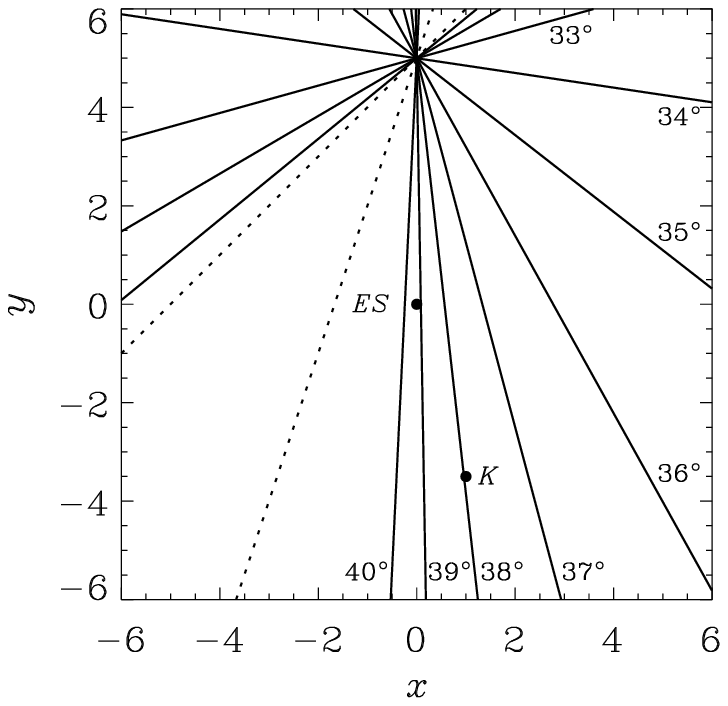]{Parameter space of opacity laws
$\kappa=\kappa_0\rho^xT^y$.  The points marked `K' and `ES' correspond
to Kramers opacity and electron-scattering opacity, respectively.
{\it Solid lines\/}: contours of the angle of inclination for which
the potential difference is minimized when expressed in physical units
(eq. [\ref{eq17}]).  The contour values are $31^\circ$, $32^\circ$,
\dots, $40^\circ$.  {\it Dotted lines\/}: contours equivalent to
$30^\circ$ and $90^\circ$; in the narrow sector between these lines, a
minimum is not obtained.\label{fig7}}

\figcaption[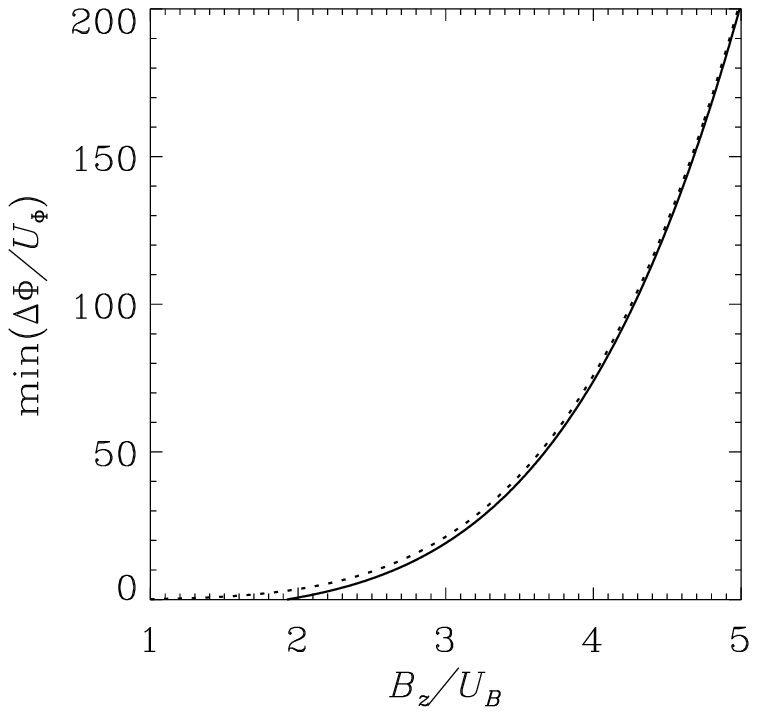]{{\it Solid line\/}: true minimum of the
potential difference with respect to $i$, as a function of the vertical
magnetic field strength, both expressed in physical units.  {\it
Dotted line\/}: the limiting form calculated from equation
(\ref{eq12}).\label{fig8}}

\pagestyle{empty}
\newpage

\centerline{\epsfbox{fig1.eps}}
\vfill
\begin{flushleft}
{\bf Figure 1}
\end{flushleft}

\newpage

\centerline{\epsfbox{fig2.eps}}
\vfill
\begin{flushleft}
{\bf Figure 2}
\end{flushleft}

\newpage

\centerline{\epsfbox{fig3.eps}}
\vfill
\begin{flushleft}
{\bf Figure 3}
\end{flushleft}

\newpage

\centerline{\epsfbox{fig4.eps}}
\vfill
\begin{flushleft}
{\bf Figure 4}
\end{flushleft}

\newpage

\centerline{\epsfbox{fig5.eps}}
\vfill
\begin{flushleft}
{\bf Figure 5}
\end{flushleft}

\newpage

\centerline{\epsfysize=18cm\epsfbox{fig6.eps}}
\vfill
\begin{flushleft}
{\bf Figure 6}
\end{flushleft}

\newpage

\centerline{\epsfbox{fig7.eps}}
\vfill
\begin{flushleft}
{\bf Figure 7}
\end{flushleft}

\newpage

\centerline{\epsfbox{fig8.eps}}
\vfill
\begin{flushleft}
{\bf Figure 8}
\end{flushleft}

\end{document}